\documentclass[%
preprint,
showpacs,preprintnumbers,
 amsmath,amssymb,
 aps,
]{revtex4-1}

\usepackage{graphicx}
\usepackage{dcolumn}
\usepackage{bm}

\begin{document}


\title{Probe of the Solar Magnetic Field \\ Using the ``Cosmic-Ray Shadow'' of the Sun}

\author{
M.~Amenomori,$^{1}$ X.~J.~Bi,$^{2}$ D.~Chen,$^{3}$ 
T.~L.~Chen,$^{4}$ W.~Y.~Chen,$^{2}$ S.~W.~Cui,$^{5}$ Danzengluobu,$^{4}$ 
L.~K.~Ding,$^{2}$ C.~F.~Feng,$^{6}$ Zhaoyang Feng,$^{2}$ Z.~Y.~Feng,$^{7}$
Q.~B.~Gou,$^{2}$ Y.~Q.~Guo,$^{2}$ K.~Hakamada,$^{8}$ H.~H.~He,$^{2}$ Z.~T.~He,$^{5}$
K.~Hibino,$^{9}$ N.~Hotta,$^{10}$ Haibing~Hu,$^{4}$ H.~B.~Hu,$^{2}$
J.~Huang,$^{2}$ H.~Y.~Jia,$^{7}$ L.~Jiang,$^{2}$ F.~Kajino,$^{11}$ K.~Kasahara,$^{12}$ 
Y.~Katayose,$^{13}$ C.~Kato,$^{14}$ K.~Kawata,$^{15}$ Labaciren,$^{4}$ G.~M.~Le,$^{2}$ 
A.~F.~Li,$^{16,6,2}$ H.~J.~Li,$^{4}$ W.~J.~Li,$^{2,7}$ C.~Liu,$^{2}$ J.~S.~Liu,$^{2}$ M.~Y.~Liu,$^{4}$ 
H.~Lu,$^{2}$ X.~R.~Meng,$^{4}$ K.~Mizutani,$^{12,17}$ K.~Munakata,$^{14}$ H.~Nanjo,$^{1}$ M.~Nishizawa,$^{18}$
M.~Ohnishi,$^{15}$ I.~Ohta,$^{19}$ H.~Onuma,$^{17}$ S.~Ozawa,$^{12}$ X.~L.~Qian,$^{6,2}$
X.~B.~Qu,$^{2}$ T.~Saito,$^{20}$ T.~Y.~Saito,$^{21}$ M.~Sakata,$^{11}$
T.~K.~Sako,$^{13}$ J.~Shao,$^{2,6}$ M.~Shibata,$^{13}$ A.~Shiomi,$^{22}$ T.~Shirai,$^{9}$
H.~Sugimoto,$^{23}$ M.~Takita,$^{15}$ Y.~H.~Tan,$^{2}$ N.~Tateyama,$^{9}$ S.~Torii,$^{12}$
H.~Tsuchiya,$^{24}$ S.~Udo,$^{9}$ H.~Wang,$^{2}$ H.~R.~Wu,$^{2}$ L.~Xue,$^{6}$ Y.~Yamamoto,$^{11}$
Z.~Yang,$^{2}$ S.~Yasue,$^{25}$ A.~F.~Yuan,$^{4}$ T.~Yuda,$^{15}$ L.~M.~Zhai,$^{2}$ H.~M.~Zhang,$^{2}$
J.~L.~Zhang,$^{2}$ X.~Y.~Zhang,$^{6}$ Y.~Zhang,$^{2}$ Yi~Zhang,$^{2}$ Ying~Zhang,$^{2}$ 
Zhaxisangzhu,$^{4}$ X.~X.~Zhou$^{7}$ \\
{\it
$^{1}$Department of Physics, Hirosaki University, Hirosaki 036-8561, Japan\\
$^{2}$Key Laboratory of Particle Astrophysics, Institute of High Energy Physics, Chinese Academy of Sciences, Beijing 100049, China\\
$^{3}$National Astronomical Observatories, Chinese Academy of Sciences, Beijing 100012, China\\
$^{4}$Department of Mathematics and Physics, Tibet University, Lhasa 850000, China\\
$^{5}$Department of Physics, Hebei Normal University, Shijiazhuang 050016, China\\
$^{6}$Department of Physics, Shandong University, Jinan 250100, China\\
$^{7}$Institute of Modern Physics, SouthWest Jiaotong University, Chengdu 610031, China\\
$^{8}$Department of Natural Science and Mathematics, Chubu University, Kasugai 487-8501, Japan\\
$^{9}$Faculty of Engineering, Kanagawa University, Yokohama 221-8686, Japan\\
$^{10}$Faculty of Education, Utsunomiya University, Utsunomiya 321-8505, Japan\\
$^{11}$Department of Physics, Konan University, Kobe 658-8501, Japan\\
$^{12}$Research Institute for Science and Engineering, Waseda University, Tokyo 169-8555, Japan\\
$^{13}$Faculty of Engineering, Yokohama National University, Yokohama 240-8501, Japan\\
$^{14}$Department of Physics, Shinshu University, Matsumoto 390-8621, Japan\\
$^{15}$Institute for Cosmic Ray Research, University of Tokyo, Kashiwa 277-8582, Japan\\
$^{16}$School of Information Science and Engineering, Shandong Agriculture University, Taian 271018, China\\
$^{17}$Saitama University, Saitama 338-8570, Japan\\
$^{18}$National Institute of Informatics, Tokyo 101-8430, Japan\\
$^{19}$Sakushin Gakuin University, Utsunomiya 321-3295, Japan\\
$^{20}$Tokyo Metropolitan College of Industrial Technology, Tokyo 116-8523, Japan\\
$^{21}$Max-Planck-Institut f\"ur Physik, M\"unchen D-80805, Deutschland\\
$^{22}$College of Industrial Technology, Nihon University, Narashino 275-8576, Japan\\
$^{23}$Shonan Institute of Technology, Fujisawa 251-8511, Japan\\
$^{24}$Japan Atomic Energy Agency, Tokai-mura 319-1195, Japan\\
$^{25}$School of General Education, Shinshu University, Matsumoto 390-8621, Japan
}
}
\collaboration{The Tibet AS$\gamma$ Collaboration}


\begin{abstract}
We report on a clear solar-cycle variation of the Sun's shadow in the 10~TeV cosmic-ray flux observed by the Tibet air shower array during a full solar cycle from 1996 to 2009.  In order to clarify the physical implications of the observed solar cycle variation, we develop numerical simulations of the Sun's shadow, using the potential field source surface (PFSS) model and the current sheet source surface (CSSS) model for the coronal magnetic field. We find that the intensity deficit in the simulated Sun's shadow is very sensitive to the coronal magnetic field structure, and the observed variation of the Sun's shadow is better reproduced by the CSSS model.  This is the first successful  attempt to evaluate the coronal magnetic field models by using the Sun's shadow observed in the TeV cosmic-ray flux.
\end{abstract}

\maketitle


\section{Introduction}
The Sun has a strong and complex magnetic field, and much of the solar activity appears to be directly connected to the properties of the magnetic field varying with a period of about 11 years.  The interplanetary magnetic field (IMF) is the term representing the solar magnetic field carried outward by the solar wind into the heliosphere as magnetic field lines from the Sun are dragged along by the highly conductive solar wind plasma \cite{Parker58}.  While the large-scale structure of the IMF is fairly simple and stable far from the Sun, the coronal magnetic field near the Sun is more complex and has not been fully understood yet.  
Since coronal magnetic fields are still difficult to observe with direct or remote measurements, they have to be extrapolated from the observed photospheric fields. A simple and widely adopted model, the potential field source surface (PFSS) model \cite{Schatten69,Hakamada95}, assumes that electric currents play a negligible role in the solar corona.  The current sheet source surface (CSSS) model, on the other hand, includes large-scale horizontal currents \cite{Bogdan86,Zhao95}.  The latter is physically more realistic and capable of reproducing the observed cusp structures in the solar corona better than the PFSS model does \cite{Schussler06}.  

The Sun with an optical diameter of about $0.5^\circ$ viewed from Earth blocks cosmic rays coming from the direction of the Sun and casts a shadow in the cosmic-ray intensity, which is possibly influenced by the solar magnetic field \cite{Clark57}.  The Tibet air shower (AS) experiment has been successfully observing the Sun's shadow at TeV energies and has confirmed, for the first time, the effect of the solar magnetic field on the shadow \cite{Amenomori93,Amenomori94}. 
In this Latter, we present the temporal variation of the Sun's shadow observed in the period of 1996 -- 2009, covering the Solar Cycle 23, and discuss the effects of the large-scale solar magnetic field by means of numerical simulations based on the coronal magnetic field models.

\section{Experiment and Data Analysis}

\begin{figure*}
\includegraphics[width=16.0cm]{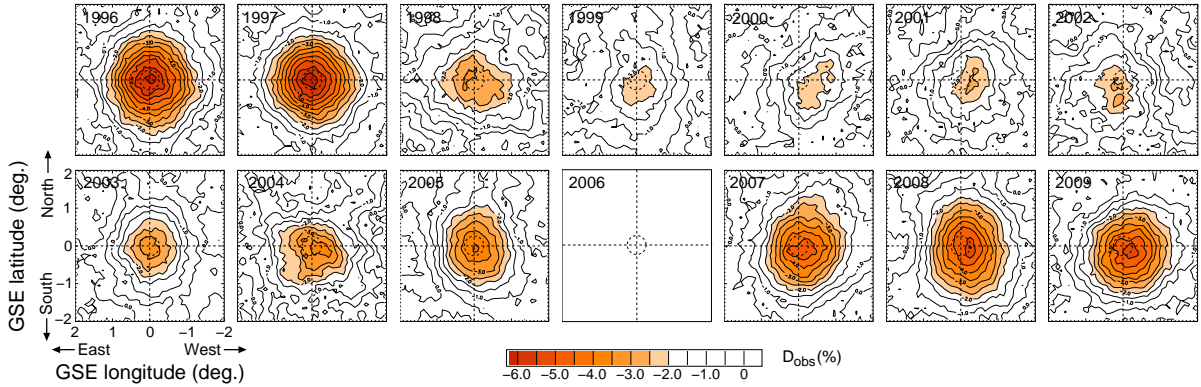}
\caption{\label{fig_1} Year-to-year variation of the observed Sun's shadow between 1996 and 2009. Each panel displays a two-dimensional contour map of the observed flux deficit ($D_{\rm obs}$).
The map in 2006 is omitted because of insufficient statistics for drawing a map. }
\end{figure*}

The Tibet AS array has been operating at Yangbajing (4,300~m above sea level) in Tibet, China since 1990. The effective area of the AS array has been gradually enlarged, in several steps, by adding 0.5~m$^2$ scintillation detectors to the preceding Tibet-I, II, and III arrays \cite{Amenomori08}. 
In this Letter, we analyze the AS events obtained by the same detector configuration as the Tibet-II array which 
started operation in 1995 \cite{Amenomori00}. 
The overall angular resolution and the modal energy of the Tibet-II array configuration are estimated to be $0.9^\circ$ and 10~TeV, respectively. 
For the analysis of the Sun's shadow, the number of on-source events ($N_{\rm on}$) is defined as the number of events arriving from the direction within a circle of $0.9^\circ$ radius centered at the given point on the celestial sphere. The number of background or off-source events ($\langle N_{\rm off} \rangle$) is then calculated by averaging the number of events within each of the eight off-source windows which are located at the same zenith angle as the on-source window \cite{Amenomori09}. 
We then estimate the flux deficit relative to the number of background events as $D_{\rm obs}=( N_{\rm on} - \langle N_{\rm off} \rangle) / \langle N_{\rm off} \rangle$ at every $0.1^\circ$ grid of Geocentric Solar Ecliptic (GSE) longitude and latitude surrounding the optical center of the Sun. 

Shown in Fig.~\ref{fig_1} are yearly maps of $D_{\rm obs}$ in $\%$ from 1996 to 2009. We exclude the year of 2006 due to low statistics.  
Inspection of Fig.~\ref{fig_1} shows that the Sun's shadow is considerably darker (with larger negative $D_{\rm obs}$) around 1996 and 2008 when solar activity was close to the minimum, while it becomes quite faint (with smaller negative $D_{\rm obs}$) around 2000 when the activity was high.

\section{MC Simulation}

\begin{figure}[h]
\includegraphics[width=6cm]{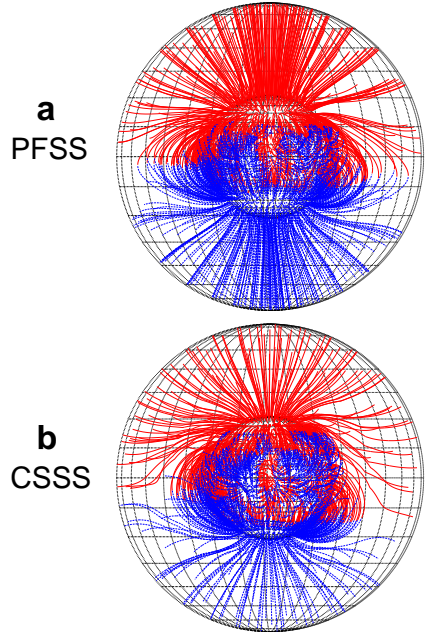}
\caption{\label{fig_2} Magnetic field line structures calculated using (a) the PFSS model and (b) the CSSS model in CR1910 (Year 1996), in a region between the photosphere and the source surface at 2.5$R_{\odot}$, represented by the inner and outer spheres, respectively. The red (blue) lines represent the field lines directing away from (toward) the photosphere.  }
\end{figure}

We have carried out Monte Carlo (MC) simulations to interpret the observed solar cycle variation of the Sun's shadow. For the primary cosmic rays, we used the energy spectra and chemical composition obtained mainly by direct observations \cite{Asakimori98,Sanuki00,Apanasenko01,Amenomori08} in the energy range from 0.3 to 1000 TeV.  We throw primary cosmic rays toward the observation site on the top of the atmosphere along the path of the Sun, and generate AS events in the atmosphere using the CORSIKA code \cite{Heck98} with the QGSJET hadronic interaction model.  These simulated AS events are fed into the detector simulation based on the Epics code \cite{Kasahara}, and are analyzed in the same way as the experimental data to deduce the AS size and the arrival direction.  An opposite charge is assigned to the primary particle of each analyzed event, and these antiparticles are shot back in random directions within a circular window of the radius of $4^\circ$ centered at the Sun from the first interaction point in the atmosphere.  A fourth order Runge-Kutta algorithm is applied to calculate the trajectory of each antiparticle in the model magnetic field described below.  We then select trajectories reaching the photosphere, and the initial shooting direction of each trajectory is tagged as a ``forbidden orbit''.  After smearing the initial shooting direction mimic the angular resolution, we finally obtain the predicted shadow.

For the coronal magnetic field, we examine two source surface (SS) models.  The SS is defined as a boundary spherical surface where magnetic field lines become purely radial, being dragged out by the supersonic solar wind.  The SS models express the magnetic field in terms of the scalar magnetic potential expanded into a spherical harmonic series which includes terms corresponding to the dipole field as well as the higher-order terms representing complicated field structures deduced from the photospheric magnetic field observations \cite{Hakamada05}.  One of the models is the PFSS model assuming current-free conditions (i.e., $\nabla \times {\bf B} \rm =0$, where ${\bf B}$ is the magnetic field vector) in the solar corona. It contains two free parameters, the radius $R_{\rm ss}$ of the SS and the order of the spherical harmonic series $n$. In this work, we set $R_{\rm ss}$ to 2.5 solar radii (2.5$R_{\odot}$) which is a realistic standard value \cite{Hakamada95}, and we set $n=10$ which is sufficient to describe fine structures relevant to the orbital motion of high energy particles.  The other model is the CSSS model \cite{Zhao95}, which includes the large-scale horizontal currents.  The CSSS model involves four free parameters, $R_{\rm ss}$, $n$, the radius $R_{\rm cp}$ ($<R_{\rm ss}$) of the spherical surface where the magnetic cusp structure in the helmet streamers appears and the length scale of horizontal electric currents in the corona $l_{a}$.  Here, we examine two different cases with $R_{\rm ss}$ = 2.5$R_{\odot}$, and $R_{\rm ss}$ = 10$R_{\odot}$.  The former is a standard value used in the original paper \cite{Zhao95} while the latter gained recent support by some evidence \cite{Zhao02,Schussler06}.  Placing the SS at 10$R_{\odot}$ or farther from the Sun yields better agreement with the latitude-independent IMF strength observed by Ulysses \cite{Balogh95} and better reproduces the solar cycle variation of the IMF magnitude observed at the Earth. 
The other parameters, $n$, $R_{\rm cp}$ and $l_{a}$ are set to 10, 1.7$R_{\odot}$, and 1.0$R_{\odot}$, respectively \cite{Schussler06}, while we find the reproduced solar cycle variation of the Sun's shadow less sensitive to these parameters.

The components of ${\bf B}$ are calculated at each point on the antiparticle's orbit in space by using the harmonic coefficients derived for every Carrington rotation (CR) period ($\sim$27.3 days) from the photospheric magnetic field observations with the spectromagnetograph of the National Solar Observatory at Kitt Peak \cite{Jones92}.  Figure~\ref{fig_2} shows the magnetic field lines obtained using (a) the PFSS model and (b) the CSSS model for the CR number 1910 (CR1910) in the solar minimum year 1996.  The CSSS model results in more field lines diverging from the polar region toward the equatorial plane than those in the PFSS model.  The radial coronal field on the SS is then stretched out to the interplanetary space forming the simple Parker-spiral IMF \cite{Parker58}.  For the radial solar wind speed we use the ``solar wind speed synoptic chart'' estimated from the interplanetary scintillation measurement in each CR and averaged over the Carrington longitude \cite{Tokumaru10,NagoyaIPS}.  In addition, we assume a stable dipole field for the geomagnetic field.


\section{Results and Discussions}

\begin{figure}
\includegraphics[width=8cm]{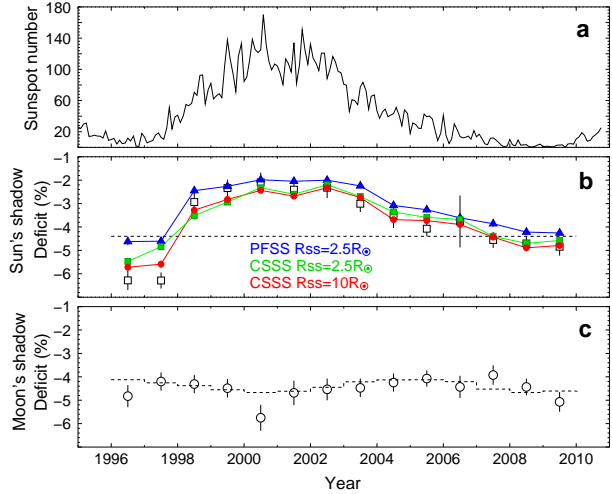}
\caption{\label{fig_3} Temporal variations of (a) the monthly mean sunspot number \cite{ngdc09b}, (b) the deficit intensity due to the Sun's shadow, and (c) the deficit intensity due to the Moon's shadow.  The open squares in the panel (b) are the observed central deficit ($D_{\rm obs}$). The blue triangles, green squares, and red circles indicate the central deficits ($D_{\rm MC}$) by the MC simulations assuming the PFSS ($R_{\rm ss}$ = 2.5$R_{\odot}$), the CSSS ($R_{\rm ss}$ = 2.5$R_{\odot}$), and the CSSS ($R_{\rm ss}$ = 10.0$R_{\odot}$) models, respectively.  The dashed lines in the panels (b) and (c) are the deficits expected from the apparent angular size of the Sun and the Moon.}
\end{figure}

\begin{table}
\caption{\label{tab1}%
Results on the $\chi^2$ test for the consistency between data and MC models using the systematic error (only the statistical error).}
\begin{ruledtabular}
\begin{tabular}{lcc}
\textrm{MC models}&
\textrm{$\chi^{2} /$ DOF}\footnote{$\chi^{2}$ is defined in Eq.~(\ref{eq_1}) and
DOF means degrees of freedom.}&
\textrm{Probability}\\
\colrule
PFSS $R_{\rm ss}$=2.5$R_{\odot}$ & 44.5(55.2)/14 & $4.9\times10^{-5}$($7.9\times10^{-7}$)\\
CSSS $R_{\rm ss}$=2.5$R_{\odot}$ & 21.1(26.2)/14 & 0.099(0.024) \\
CSSS $R_{\rm ss}$=10$R_{\odot}$ & 8.3(10.3)/14 & 0.87(0.74)\\
\end{tabular}
\end{ruledtabular}
\end{table}

For quantitative analysis of the temporal variation of the Sun's shadow, we use the central deficit, $D_{\rm obs}$, measured at the center of the two-dimensional map in Fig.~\ref{fig_1}.  Open squares in Fig.~\ref{fig_3}(b) indicate the temporal variation of this central $D_{\rm obs}$.  One can see that the magnitude of the central $D_{\rm obs}$ in panel (b)  varies in a clear anticorrelation with the sunspot number in Fig.~\ref{fig_3}(a) \cite{ngdc09b}.  The amplitude of the yearly variation of the central $D_{\rm obs}$ is as large as 50\% of the deficit expected from the apparent angular size of the Sun shown by the horizontal dashed line. This is remarkably different from variation of the Moon's shadow in Fig.~\ref{fig_3}(c), which remains stable during the whole period (see \cite{Amenomori09} for more details about the Moon's shadow).  
Since the Moon and the Sun observed from Earth have almost the same angular size, the stable deficit due to the Moon's shadow provides a good estimate of any conceivable systematic or instrumental effect.  The dashed line in Fig.~\ref{fig_3}(c) indicates the deficit expected from the Moon's apparent angular size, which undergoes an approximately $\pm$0.26\% variation due to the variation of the distance between the Earth and the Moon.  We estimate that the observed deficit due to the Moon's shadow, averaged from 1996 to 2009, is $\langle D^{\rm moon}_{\rm obs} \rangle \pm \langle \sigma^{\rm moon}_{\rm obs} \rangle = (-4.46\pm0.11)\%$, while for the expected one we find $\langle D^{\rm moon}_{\rm exp} \rangle = -4.30\%$.  From this, the systematic error for the absolute deficit is estimated to be $\langle \sigma_{\rm sys} \rangle = \sqrt{(\langle D^{\rm moon}_{\rm obs} \rangle - \langle D^{\rm moon}_{\rm exp} \rangle )^2 + \langle \sigma^{\rm moon}_{\rm obs} \rangle^2} = 0.19\% $.

From the MC simulations of the Sun's shadow, we calculate the central deficit ($D_{\rm MC}$) equivalent to $D_{\rm obs}$ by $D_{\rm MC} = -N_{\rm hit} / N_{\rm all}$ for each coronal field model where $N_{\rm all}$ is the number of all initial shooting directions within the 0.9$^\circ$ circle centered at the Sun and $N_{\rm hit}$ is the number of events hitting the Sun.  The blue triangles, green squares and red circles show $D_{\rm MC}$ assuming the PFSS ($R_{\rm ss}$ = 2.5$R_{\odot}$), the CSSS ($R_{\rm ss}$ = 2.5$R_{\odot}$) and the CSSS ($R_{\rm ss}$ = 10.0$R_{\odot}$) models, respectively.  For quantitative comparisons between observations and the MC expectations in Fig.~\ref{fig_3}(b), we perform a $\chi^{2}$ test as
\begin{equation}
\chi^{2} = \sum_{i=1}^{14} \frac{\left(D^{i}_{\rm obs} - D^{i}_{\rm MC}
\right)^2}{(\sigma^{i}_{\rm obs})^2 + (\sigma^{i}_{\rm MC})^2 + \langle 
\sigma_{\rm sys} \rangle^2},
\label{eq_1}
\end{equation}
where $D^{i}_{\rm obs}$ and $D^{i}_{\rm MC}$ are the observed and predicted central deficits for the {\it i}th year from 1996, while $\sigma^{i}_{\rm obs}$ and $\sigma^{i}_{\rm MC}$ are their respective (statistical) errors.  The results of the $\chi^{2}$ test are summarized in Table~\ref{tab1}.  For the PFSS model, the $\chi^2$ yields a very low likelihood of $4.9\times10^{-5}$ since the MC simulations assuming the PFSS model yield too small $D_{\rm MC}$ to explain the observed deficit shown in Fig.~\ref{fig_3}(b).  The difference is particularly significant in 1996 and 1997.
On the other hand, the predictions assuming the CSSS model (green squares) are in good agreement with the observations.  Furthermore, we also find that the CSSS model with $R_{\rm ss}$ = 10.0$R_{\odot}$ shown by red circles gives, overall, an even better agreement than the case with $R_{\rm ss}$ = 2.5$R_{\odot}$.  
We note that the PFSS model assuming $R_{\rm ss} \gg 2.5R_{\odot}$ was omitted from simulations, because this magnetic field is dominated by the unrealistic closed field lines.

\begin{figure}
\includegraphics[width=8cm]{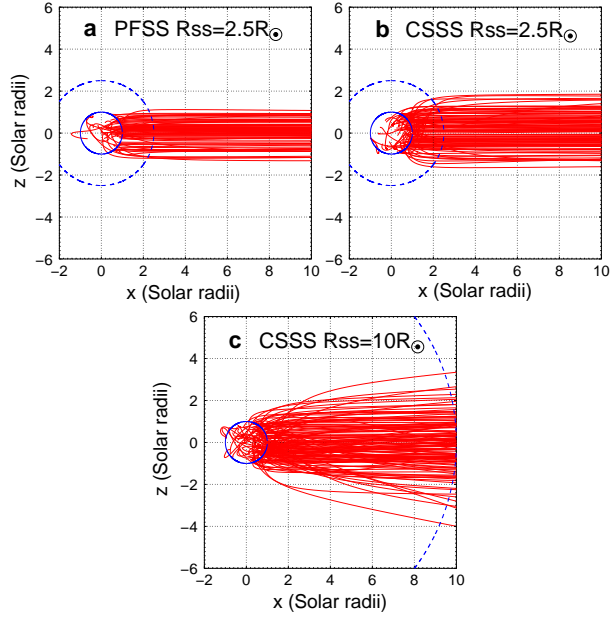}
\caption{\label{fig_4}
Simulated trajectories of antiparticles ejected toward the Sun from the Earth in CR1910 (year 1996) presented in HEE coordinates. Only trajectories of antiparticles hitting the Sun are plotted.  The three panels refer to simulations assuming (a) the PFSS ($R_{\rm ss}$ = 2.5$R_{\odot}$), (b) the CSSS ($R_{\rm ss}$ = 2.5$R_{\odot}$) and (c) the CSSS ($R_{\rm ss}$ = 10.0$R_{\odot}$) model, respectively.  The inner solid and outer dashed circles indicate the size of the photosphere and the SS, respectively.}
\end{figure}

The model dependence of the predicted deficits in the MC simulation can be interpreted in terms of the cosmic-ray trajectories in the different magnetic field models.  Figure~\ref{fig_4} shows sample trajectories in heliocentric earth ecliptic (HEE) coordinates of antiparticles hitting the Sun in CR1910 (Year 1996) in three cases we consider : (a) the PFSS ($R_{\rm ss}$ = 2.5$R_{\odot}$), (b) the CSSS ($R_{\rm ss}$ = 2.5$R_{\odot}$) and (c) the CSSS ($R_{\rm ss}$ = 10.0$R_{\odot}$) model.  The same number of antiparticles are ejected from Earth in each simulation with the modal energy of 10 TeV similarly to analyzed AS events.  The trajectories assuming the PFSS model are almost straight lines inside the SS, while the trajectories assuming the CSSS model are strongly deflected in the polar region at high latitudes.  As shown in Fig.~\ref{fig_2}, the polar field lines of the CSSS model stretch out to lower latitudes, and they face more toward the Earth just inside the SS.  This becomes more visible when the SS is set to farther from the Sun at $R_{\rm ss}$ = 10.0$R_{\odot}$.  Hence, antiparticles can easily move along the open field lines through the SS and reach toward polar region on the photosphere.  This focusing effect results in a larger deficit than the one expected from the Sun's apparent angular size.  At the solar maximum, on the other hand, the Sun's shadow diminishes due to the antiparticles' orbits being deflected in the complicated and disordered coronal field and excluded from hitting the photosphere.

The Sun's shadow observed by the Tibet AS array offers a powerful tool for analyzing the solar magnetic field quantitatively. It is noted, however, that building a unique coronal magnetic field only from the observation of the Sun's shadow would be difficult, since the observed Sun's shadow reflects not only the coronal magnetic field, but also the integrated IMF between the Sun and the Earth.
We conclude that the Sun's shadow is better reproduced by the CSSS model than by the PFSS model. We find that the flux deficit in the Sun's simulated shadow is very sensitive to the coronal magnetic field structure, which is still difficult to observe with direct or remote measurements.  This is the first successful attempt to evaluate the coronal field models by using the Sun's shadow observed in TeV cosmic rays.  


\begin{acknowledgments}
The collaborative experiment of the Tibet Air Shower Arrays has been conducted under the auspices of the Ministry of Science and Technology of China and the Ministry of Foreign Affairs of Japan. This work was supported in part by a Grant-in-Aid for Scientific Research on Priority Areas from the Ministry of Education, Culture, Sports, Science and Technology, by Grants-in-Aid for Science Research from the Japan Society for the Promotion of Science in Japan, and by the Grants from the National Natural Science Foundation of China and the Chinese Academy of Sciences. The authors thank Dr. Xuepu Zhao of Stanford University for providing the usage of the CSSS model. The authors thank Dr. J\'{o}zsef K\'{o}ta of University of the Arizona for his useful comments and discussions.
\end{acknowledgments}

\nocite{*}

\bibliography{ms}

\end{document}